# Determining the upper critical magnetic field for N-doped lutetium hydride directly from the source data files in Dasenbrock-Gammon *et al.*, Nature **615**, 244 (2023)

Dale R. Harshman [1,*] and Anthony T. Fiory [2]

[1] *Physikon Research Corporation, Lynden, WA 98264, USA*
[2] *Bell Labs Retired, Summit, NJ 07901, USA*
(19 May 2023)

The Ginzburg-Landau-based upper critical magnetic field $H_{C2}(0) \approx 88$ T for N-doped lutetium hydride, reported in Dasenbrock-Gammon *et al.*, Nature **615**, 244 (2023), is obtained therein by modeling resistance behavior, defining transitions widths, and applying magnetic fields $H = 1$ T and 3 T. A method is presented herein for determining the critical temperature $T_C(H)$ directly from the resistance drops in the source data, implying a temperature slope $-dH_{C2}/dT$ of 0.46(6)–0.51(5) T/K and, by applying pure BCS theory, an $H_{C2}(0)$ of 71(10)–79(8) T.

---

* drh@physikon.net

Apart from values of the transition temperature $T_C$ measured at various applied pressures, the results reported for nitrogen-doped lutetium hydride in Dasenbrock-Gammon *et al.* also provide quantitative information on the zero-temperature upper critical magnetic field, calculated therein to be $H_{C2}(0) \approx 88$ T at pressure $P = 15$ kbar ($P \sim 10$ kbar is optimal) [1]. Based on the Ginzburg-Landau model, this result is obtained from electrical resistance behavior under applied magnetic field $H$ (0 T, 1 T, and 3 T) and pressure (15 kbar), which is displayed as three normalized relative resistance curves in Extended Data Fig. 15 [1]. The authors' data analysis, as described in the figure caption, determines $H_{C2}$ from increases in defined superconducting transition widths under external magnetic fields. The displayed curves show drops from near unity at high temperatures to nearly zero in low-temperature regions. The source resistance-*vs.*-temperature data, available on-line [2], show drops in resistance, decreasing nearly monotonically by ~80% at low temperature [2]. Side-by-side figures showing the temperature dependence of both the normalized relative resistance and source resistance may be viewed, e.g., in Ref. 3. Of particular interest, as noted in Ref. 3, is the large amount of inhomogeneous variation among the samples measured in Ref. 1 as indicated in the variation in the transition widths. While one may plausibly conclude an absence of superconductivity [3], this variation may also reflect sample growth and measurement difficulties; the stated success rate of measuring a sample with superconducting properties is only about 35% [1].

Proceeding under the assumption that the resistance drops denote superconducting transitions of type II superconducting condensates contained within the sample, the transition temperature $T_C(H)$ in an applied magnetic field $H$ measures the upper critical field as $H_{C2}(T_C(H)) = H$ at the temperature $T_C(H)$.

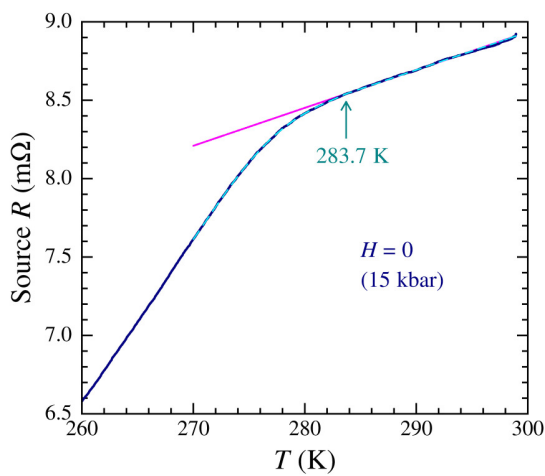

FIG. 1. Source resistance data for $H = 0$ T from Ref. [2] for N-doped lutetium hydride at $P = 15$ kbar. Dashed cyan overlay curve and solid magenta line denote functions $F$ and $L$, respectively, in Eq. (1). Marked arrow indicates $T_{\text{onset}}$.

In the source data for $H = 0$ T, 1 T, and 3 T at $P = 15$ kbar, the resistance transitions are rounded by presumed inhomogeneity in addition to thermodynamic fluctuations. However, it remains feasible to derive fairly accurate estimates of $T_C$ by modeling a functional form for the experimental superconducting transition. Resistance varies nearly linearly above the transition and drops below linear at an onset temperature $T_{\text{onset}}$. This form is modeled by the following function,

$$F(T) = L(T - T_{\text{onset}}) - \theta(T_{\text{onset}} - T)\, P(T - T_{\text{onset}}), \quad (1)$$

where $L(x) = r_0 + r_1 x$ and $P(x) = \sum_{n=1}^{4} a^n x^n$ are linear and polynomial functions of $x = T - T_{\text{onset}}$, respectively; $\theta(-x)$ is the unit step function. Results for fitting Eq. (1) to the source data for $H = 0$ T in the transition region $T_1 \leq T \leq T_2$ ($T_1 = 270$ K, $T_2 = 299$ K) is shown in Fig. 1, using a 4th order polynomial for $P$. Data are shown in dark blue; $F$ is the dashed cyan curve overlaying the data, and the linear function $L$ is shown by the magenta line. The fit yields $T_{onset} = 283.7$ K with statistical uncertainty under 0.1 K. The Appendix (see Fig. 11) checks this method against the near-optimal $P = 10$ kbar data with onset $T_C = 294$ K [1].

The resistance drop is thus quantified by the deviation from the linear function $L$, as shown in Fig. 2. An estimate for the transition temperature is the point where the resistance drop is readily distinguishable from noise, taken here to be $3\sigma$, where $\sigma = 0.0061$ m$\Omega$ is the rms deviation for $T > T_{onset}$. This point, denoted as $T_{3\sigma}$, is at the intersection of the data and the dashed line, as marked by the magenta arrow, and provides 99.7 % confidence that $T_C\,(H=0)$ is at least 281.5 K at $P = 15$ kbar.

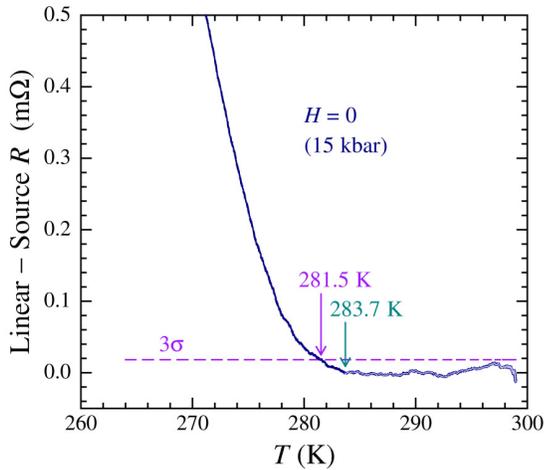

FIG. 2. Difference between the linear function $L$ and source resistance for N-doped lutetium hydride at $P = 15$ kbar; the region $T > T_{onset}$ is overlaid in a lighter tone. Dashed line denotes 3 standard deviations above zero. The marked cyan and magenta arrows denote $T_{onset}$ and $T_{3\sigma}$, respectively.

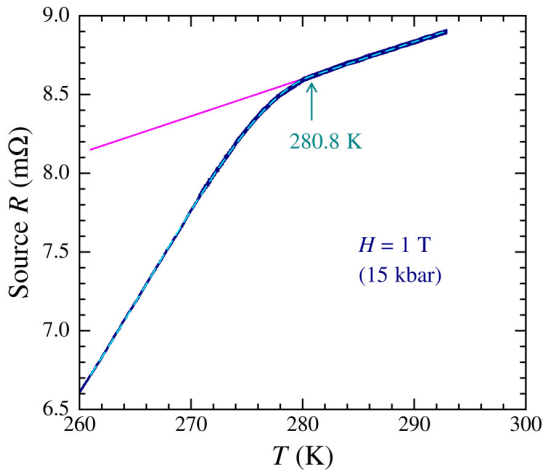

FIG. 3. Analysis of source resistance under $H = 1$ T.

Figures 3 through 6 show results of analyses conducted in a similar manner for source resistance data under applied magnetic fields. Parameters are in Table 1.

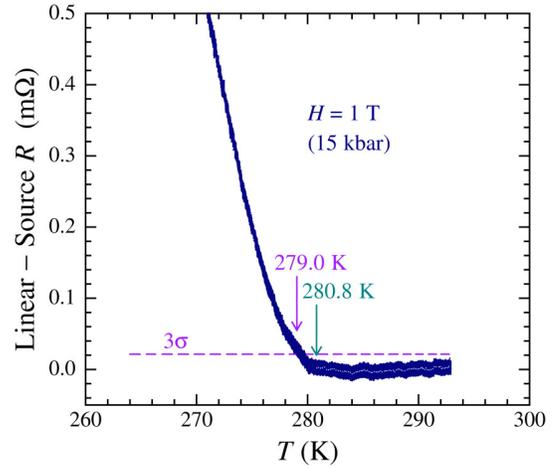

FIG. 4. Analysis of source resistance under $H = 1$ T.

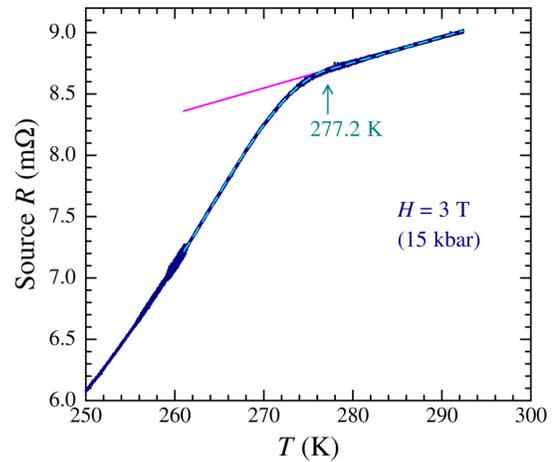

FIG. 5. Analysis of source resistance under $H = 3$ T.

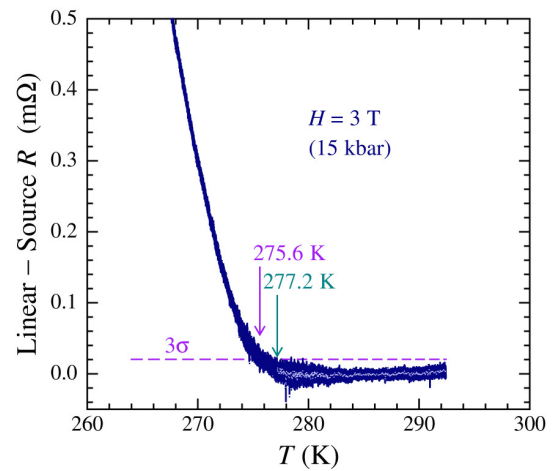

FIG. 6. Analysis of source resistance under $H = 3$ T.

Table 1. Parameters obtained by fitting Eq. (1) to the source resistance data for $P = 15$ kbar from Ref. [2].

| $H$ (T) | $T_{3\sigma}$ (K) | $T_{onset}$ (K) | $r_0$ (mΩ) | $r_1$ (mΩK$^{-1}$) | $a_1$ (mΩK$^{-1}$) | $a_2$ (mΩK$^{-2}$) | $a_3$ (mΩK$^{-3}$) | $a_4$ (mΩK$^{-4}$) | $\sigma$ (mΩ) | $T_1$ (K) | $T_2$ (K) | rms fit (mΩ) |
|---|---|---|---|---|---|---|---|---|---|---|---|---|
| 0 | 281.5 | 283.7 | 8.541 | 0.0242 | –0.00832 | –0.00160 | –0.000616 | –2.27E-5 | 0.00605 | 270 | 298.9 | 0.0054 |
| 1 | 279.0 | 280.8 | 8.617 | 0.0236 | 0.000274 | 0.00735 | 0.000224 | 1.92E-6 | 0.00711 | 260 | 292.4 | 0.0065 |
| 3 | 275.6 | 277.2 | 8.697 | 0.0208 | –0.000202 | 0.00648 | 6.48E-5 | –4.12E-6 | 0.00675 | 260 | 289.6 | 0.0095 |

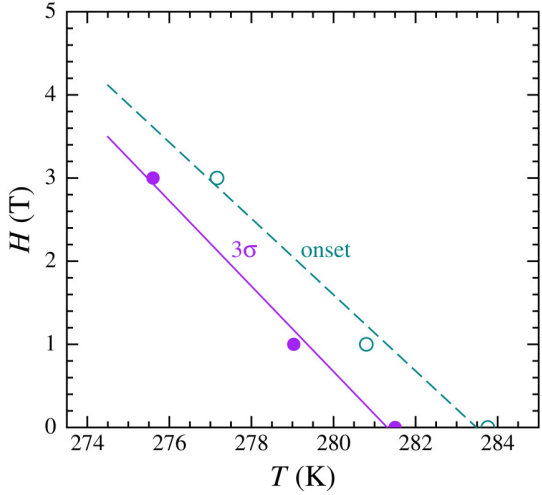

FIG. 7. Variation of applied field $H$ with temperatures $T = T_{3\sigma}$ (solid symbols) and $T = T_{onset}$ (open symbols) and corresponding linear fits given by solid and dashed lines, respectively. $P = 15$ kbar.

The source resistance data under applied magnetic fields contain oscillations in the vicinity of the digitization Nyquist frequency, as indicated by the banded scatter of points in the difference plots of Figs. 4 and 6 (details in Appendix). For clarity, the overlays for $T > T_{onset}$ in these figures were obtained by compressing the number of data points by a factor of 4, which partly suppresses the oscillations.

Variations of $H$ with $T_{3\sigma}$ and $T_{onset}$ are shown as the solid and open symbols, respectively, in Fig. 7. The solid line is a linear regression fit of $H$ vs. $T_{3\sigma}$, yielding a slope of –0.51(5) T/K and an intercept of 281.3(3) K; the dashed line is a fit of $H$ vs. $T_{onset}$ with slope –0.46(6) T/K and intercept 283.5(5) K. Considering these results as estimates of $H$ vs. $T_C(H)$, one may equate $dH_{C2}/dT$ near $T_C$ to the slopes and $T_C$ to the intercepts. The expression $H_{C2} = \varphi_0/2\pi\xi^2$, where $\varphi_0$ is the flux quantum, coupled with the pure BCS limit near $T_C$ for the temperature dependence of the coherence distance $\xi = 0.74\,\xi_0\,(1 - T/T_C)^{-1/2}$ [4], then gives $\xi_0$ between 20(1) and 21(1) Å, and $H_{C2}(0)$ (= $\varphi_0/2\pi\xi_0^2$) ranging from 79(8) to 71(10) T, reflecting the variations of $H$ with $T_{3\sigma}$ and $T_{onset}$, respectively.

## CONCLUSION

Analysis of resistance drops determined from the source data of Extended Data Fig. 15 in Ref. [1] yield an $H_{C2}(0)$ of 71(10) – 79(8) T at $P = 15$ kbar. The extracted $H$-vs.-$T$ behavior exhibited in Fig. 7 is typical of a type II superconductor, which directly conflicts with the notion of an absence of superconductivity as suggested in Ref. 3.

## ACKNOWLEDGEMENT

The authors are grateful for support from the University of Notre Dame.

## APPENDIX

Figures 8 – 10 for $H = 0$, 1 T, and 3 T, respectively, show the differences between the source resistance and the fitted functions $F$ in regions near the transitions. Index numbers count sequential data points as indicated in the captions. The upper abscissa labels the corresponding temperature ranges.

Figure 11 shows the source data [5] for $P = 10$ kbar (highest onset $T_C = 294$ K reported in Fig. 2 of Ref. [1]) and a fit of Eq. (1), yielding $T_{onset} = 294.8$ K and $T_{3\sigma} = 294.6$ K. Fitting parameters are given in Table 2.

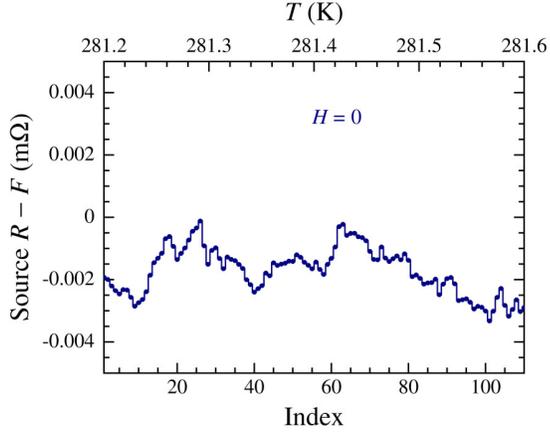

FIG. 8. Difference between the source resistance for $H = 0$ and fitted function $F$ in a temperature region near the transition. Lower abscissa denotes relative index number, upper abscissa shows temperature. Lines connecting symbols are guide to the eye. Index = 1 is data point No. 22835 in Ref. [2].

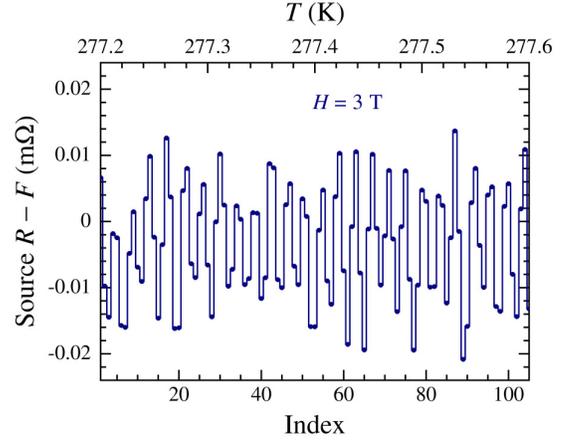

FIG. 10. Difference between the source resistance for $H = 3$ T and fitted function $F$ in a temperature region near the transition. Index = 1 is data point No. 23449.

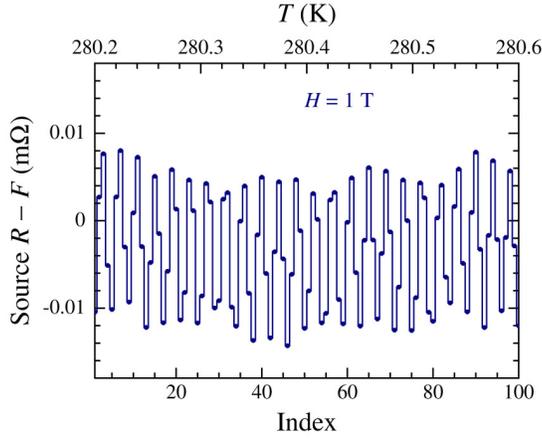

FIG. 9. Difference between the source resistance for $H = 1$ T and fitted function $F$ in a temperature region near the transition. Index = 1 is data point No. 32046.

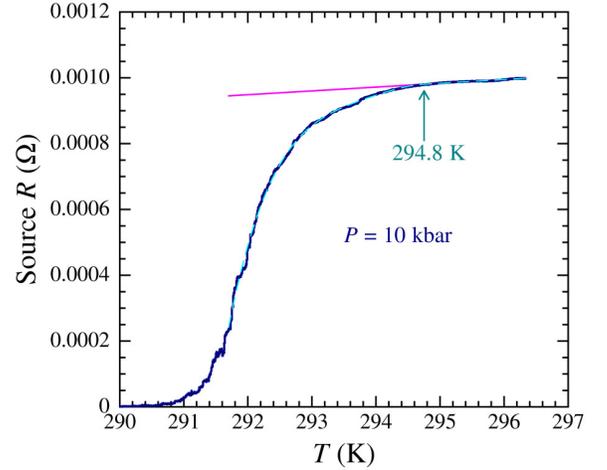

FIG. 11. Source resistance data at $P = 10$ kbar from Ref. [5]. Dashed cyan overlay curve and solid magenta line denote functions $F$ and $L$, respectively, in Eq. (1). Marked arrow indicates $T_{\text{onset}}$.

Table 2. Parameters obtained by fitting Eq. (1) to the source resistance data for $P = 10$ kbar [1, 5].

| $T_{3\sigma}$ (K) | $T_{\text{onset}}$ (K) | $r_0$ (m$\Omega$) | $r_1$ (m$\Omega$K$^{-1}$) | $a_1$ (m$\Omega$K$^{-1}$) | $a_2$ (m$\Omega$K$^{-2}$) | $a_3$ (m$\Omega$K$^{-3}$) | $a_4$ (m$\Omega$K$^{-4}$) | $\sigma$ (m$\Omega$) | $T_1$ (K) | $T_2$ (K) | rms fit (m$\Omega$) |
|---|---|---|---|---|---|---|---|---|---|---|---|
| 294.6 | 294.8 | 0.9804 | 0.01155 | 0.003200 | 0.08071 | 0.05776 | 0.01849 | 0.00108 | 291.7 | 296.3 | 0.0048 |